\newcommand{\Ks}{K$_{s}$}
\newcommand{\fout}{$f_\mathrm{out}$}
\newcommand{\Gmk}{$(G-K_s)$}
\newcommand{\MK}{$\mathrm{M}_{\mathrm{K_s}}$}

\newcommand{\acronym}[1]{{\small{#1}}}

\newcommand{\project}[1]{\textsl{#1}}
\newcommand{\gaia}{\project{Gaia}}
\newcommand{\WISE}{\project{WISE}}
\newcommand{\tmass}{\project{2MASS}}

\newcommand{\hipparcos}{\project{Hipparcos}}

\newcommand{\tgas}{\project{\acronym{TGAS}}}
\newcommand{\hmodel}{hierarchical model}

\documentclass[useAMS,usenatbib,letter]{mn2e}
\usepackage{graphicx}
\usepackage{amssymb}
\usepackage{amsmath}
\usepackage{times}
\usepackage{subfigure}
\usepackage[T1]{fontenc}
\usepackage{aecompl}
\usepackage[breaklinks,colorlinks,citecolor=blue]{hyperref}

\begin{document}

\title[Calibrating Red Clump stars with Gaia]{Red clump stars and Gaia: Calibration of the standard candle using a hierarchical probabilistic model}
 \author[K.~Hawkins et. al.]{Keith~Hawkins$^{1}$\thanks{E-mail: khawkins@astro.columbia.edu \newline Simons Fellow}, Boris~Leistedt$^{2}$\thanks{NASA Einstein Fellow}, Jo~Bovy$^{3,4}$\thanks{Alfred P. Sloan Fellow}, David~W.~Hogg$^{2,3}$,  \\ 
$^{1}$Department of Astronomy, Columbia University, 550 W 120th St, New York, NY 10027, USA \\
$^{2}$Center for Cosmology and Particle Physics, Department of Physics, New York University, 726 Broadway, New York, NY 10003, USA\\
$^{3}$Center for Computational Astrophysics, Flatiron Institute, 162 5th Ave, New York, NY 10010, USA \\
 $^{4}$Department of Astronomy and Astrophysics, University of Toronto, 50 St. George Street, Toronto, ON M5S 3H4, Canada
 }


\date{Accepted .... Received ...; in original form ...}


\maketitle

\label{firstpage}

\begin{abstract}
Distances to individual stars in our own Galaxy are critical in order to piece together the nature of its velocity and spatial structure. Core helium burning red clump (RC) stars have similar luminosities, are abundant throughout the Galaxy, and thus constitute good standard candles. We build a hierarchical probabilistic model to quantify the quality of RC stars as standard candles using parallax measurements from the first Gaia data release. A unique aspect of our methodology is to fully account for (and marginalize over) parallax, photometry, and dust corrections uncertainties, which leads to more robust results than standard approaches. We determine the absolute magnitude and intrinsic dispersion of the RC in \tmass\ bands J, H, \Ks, \gaia\ G band, and \WISE\ bands W1, W2, W3, and W4. We find that the absolute magnitude of the RC is $-1.61 \pm 0.01$ (in \Ks), $+0.44 \pm 0.01$ (in G) , $-0.93 \pm 0.01$ (in J), $-1.46 \pm 0.01$ (in H), $-1.68 \pm 0.02$ (in W1),  $-1.69 \pm 0.02$ (in W2),  $-1.67 \pm 0.02$ (in W3),  $-1.76 \pm 0.01$~mag (in W4). The mean intrinsic dispersion is $\sim$0.17 $\pm$ 0.03 mag across all bands (yielding a typical distance precision of $\sim$8\%). Thus RC stars are reliable and precise standard candles. In addition, we have also re-calibrated the zero-point of the absolute magnitude of the RC in each band, which provide a benchmark for future studies to estimate distances to RC stars. Finally, the parallax error shrinkage in the hierarchical model outlined in this work can be used to obtain more precise parallaxes than Gaia for the most distant RC stars across the Galaxy.

\end{abstract}

\begin{keywords}
Stars: distances, Stars: fundamental parameters, Stars: statistics
\end{keywords}

\section{Introduction}
Estimating distances to individual stars is a difficult undertaking and yet it is critical to understand the spatial and dynamical nature of our Galaxy. One approach to derive distances to individual stars has been to use red clump (RC) stars \citep[e.g.][]{Paczyski1998, Stanek1998, Udalski2000, Alves2002, Laney2012, Bovy2014}. The RC is a visually striking feature in the color-magnitude diagram defined by a group of evolved stars which have undergone the helium flash and all have roughly a single absolute magnitude, making the RC stars a so-called 'standard candle'. Thus their apparent magnitude is directly related to the distance of the star after accounting for extinction. 

The RC has largely been used to determine the distances to nearby galaxies \citep[e.g.][]{Stanek1998, Udalski1998, Laney2012} or stars within our own Galaxy \citep[e.g.][]{Paczyski1998, McWilliam2010, Bovy2014, Nidever2014}. However, in order to do this, three ingredients are needed: (1) a reference value (zero-point) of the absolute magnitude of the RC, (2) the interstellar extinction to the RC population of interest, and (3) a slight correction for the population effect between the local sample of RC stars used to derive the reference value and the actual RC population of interest \citep[e.g.][]{Girardi2001, Bovy2014, Girardi2016}. This work will focus on the first two. Ultimately, the intrinsic absolute magnitude of the RC can be used to not only to estimate the distance to stars across the Galaxy but also in other galaxies \citep[e.g.][]{Laney2012}, where \gaia\ will not be able to measure precise parallaxes.  

The first data release from the \gaia\ mission has produced precise parallaxes and apparent magnitudes for more than 2.5 million sources \citep{Michalik2015,Gaia2016} in the Milky Way, presenting an exciting opportunity to test the quality of using the RC as a standard candle and update the reference absolute magnitude in many bandpasses in a homogenous way. Therefore, in this Letter, we aim to use data from the astrometric \gaia\ spacecraft in combination with a collection of RC samples containing more than 970 RC stars, to measure the mean and dispersion in the absolute magnitude of the RC in several bands including \tmass\ J, H, \Ks, \WISE\ W1, W2, and W3, and \gaia\ G. This represents one of the largest samples (and one of the first with Gaia data) of RC stars used to make such a measurement.  

To achieve these aims, this Letter is organized in the following way: In section \ref{sec:data} we describe the four RC data samples that are used in this study along with the combined sample. The hierarchical model that we use to estimate the mean and dispersion in the absolute magnitude of the RC is described in section \ref{sec:method}. We present the calibration of the absolute magnitude of the RC using \gaia\  and discuss our results in the context of the literature in section \ref{sec:result}. Finally, we discuss the status of using RC as standard candle in section~\ref{sec:summary}.

\section{Data} \label{sec:data}
In this section, we describe the various RC samples that have been sourced from the literature and combined into a master sample. Ultimately, our aim is to derive an intrinsic absolute magnitude and dispersion in \tmass\ J, H, \Ks, \WISE\ W1, W2, W3, and W4, and \gaia\ G bands for the RC. To do this, we source the apparent magnitude and their uncertainties for each star from the \tmass\ survey \citep[J, H, and \Ks,][]{Cutri2003}, the \WISE\ survey \citep[W1, W2, W3, and W4,][]{Wright2010}, and first data release of the \gaia\ survey \citep[G,][]{Gaia2016,vanLeeuwen2017}. The measured trigonometric parallaxes ($\hat{\varpi}_{i}$) and their uncertainties for each star are obtained from the Tycho-Gaia Astrometric Solution \citep[\tgas,][]{Michalik2015,Gaia2016,Lindegren2016}. The median uncertainties in our RC sample are 0.02, 0.03, 0.02, 0.01, 0.01, 0.02, 0.02, 0.12~mag in the J, H, \Ks, G, W1, W2, W3, and W4 bands and 0.27~mas in $\varpi$ ($\sim$18\%). We note that for some of the brightest stars in our sample, the 2MASS photometry is saturated and thus have poor photometric quality flags. These stars also have significantly larger photometric uncertainties which are fully accounted for in our \hmodel. As such, removing them does not significantly affect our inferred absolute magnitude.

Each star's apparent magnitude and parallax and their associated uncertainties represents the observed data, $\mathcal{D}_i$, that we use to constrain the parameters of our \hmodel\ in section~\ref{sec:method}. In addition to these, an initial reddening value, E$(B-V)_{i}$, for each star is taken from the 3-dimensional (3D) dustmap of \cite{Green2015}, assuming the distance to the star is 1/$\varpi$\footnote{We note here that we do not recompute the E$(B-V)_{i}$ value from the 3D dustmap on for each chain. We instead use the first estimate as a weakly informative prior.}. We note here that this is only used to place a prior on the inferred extinction posterior. The reddening is converted into a band-specific extinction by multiplying E$(B-V)_{i}$ by an extinction coefficient which is taken from several literature sources. The extinction coefficients for J, H, K, W1, and W2 are taken from Table 2 of \cite{Yuan2013}, while the coefficients for W3 and W4 were taken from Table 4 \cite{Xue2016} assuming that $\frac{\mathrm{A}_{\mathrm{J}}}{\mathrm{A}_{\mathrm{K_s}}}$~=~2.72. Finally, the coefficient for \gaia\ G-band was derived using the information in Table 13 of \cite{Jordi2010} for stellar parameters consistent with the mean RC population. All of these coefficients can be found in the last column of Table~\ref{tab:results}.

\begin{figure}
	 \includegraphics[width=\columnwidth]{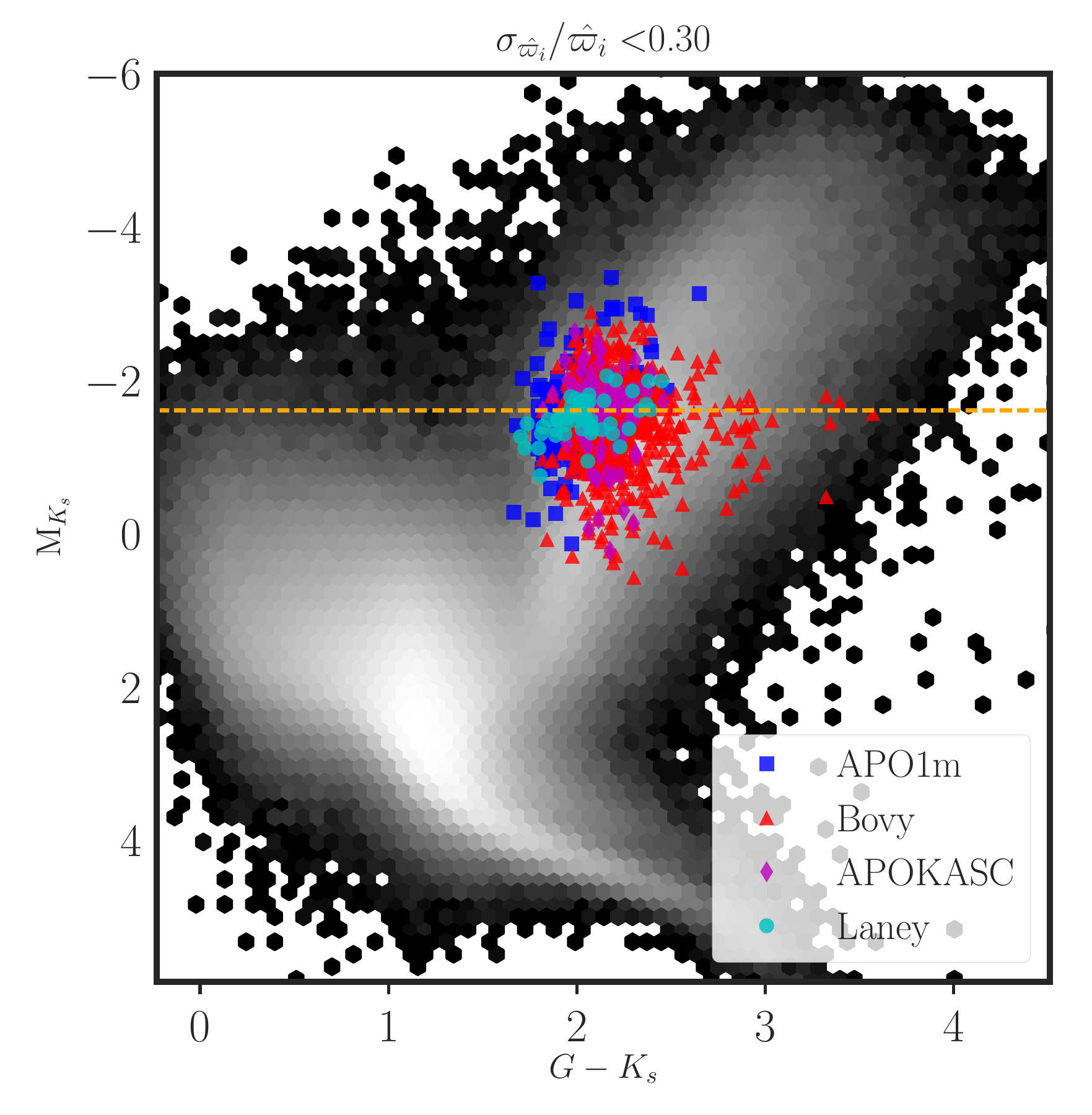}
	\caption{The naive absolute \Ks-band magnitude as a function of \Gmk\ for the \gaia-\tmass\ overlap (gray log density scale), and the red clump samples taken from APOKASC \citep[magenta diamonds,][]{Elsworth2016}, APO1m \citep[blue square,][]{Feuillet2016} APOGEE \citep[red triangles,][]{Bovy2014}, and Laney \citep[cyan circles,][]{Laney2012}. The orange dotted line is the calibrated absolute magnitude of the red clump in \MK\ from \protect\cite{Laney2012}. In all cases the parallax precision is better than 30\%.}
	\label{fig:CMD_RCall}
\end{figure}

For all of the samples described below, we choose only those RC stars with parallax uncertainties less than 30\%. We don't consider stars with noisier parallaxes because the data does not bring significant information in constraining the \hmodel\ and the posterior belief about their distances is mostly driven by the distance prior adopted rather than the data itself  \citep[e.g.][]{Bailer-Jones2015,Astraatmadja2016}. The final sample contains 972  red clump stars. We note here that the red giant branch (RGB) does overlap with the RC population thus contamination is likely. For this reason, our model will have an extra broad component to describe outlier objects, unlikely to be RC stars. In the below subsections, we describe each subsample from which we built our `master' sample and the cuts that were used. In Figure~\ref{fig:CMD_RCall} we show the \Gmk\ color-magnitude digram for probably RC stars (color symbols) in all samples with and the full \tgas-\tmass\ overlap with parallax uncertainties lower than 30\%. As this figure is for illustration purposes, to obtain the absolute magnitudes for all samples we used the distance $r_i = 1/\hat{\varpi}_i$ where $\hat{\varpi}_i$ is the \tgas\ parallax estimate. 

\subsection{APOGEE Red Clump Catalogue and APO1m Samples} \label{subsec:APOGEEsamp}
Many of the RC stars used in this work are taken from the Apache Point Observatory Galactic Evolution Experiment (APOGEE).  The APOGEE survey is a public near-infrared H-band high-resolution (R$\sim$22,500) spectroscopic survey within SDSS-IV \citep{Majewski2015}. Most spectra from the APOGEE survey have relatively high signal-to-noise ratios allowing for precise determinations of stellar parameters and chemical abundance  \citep[e.g.][]{Holtzman2015, Ness2015,Hawkins2016b}. These stellar parameters are used, in combination with PARSEC isochrones, to develop a sample of possible RC stars \citep[we refer the reader to section 2.2 of][for a detailed discussion on the selection of this sample]{Bovy2014}. We note that it is expected that the RGB contamination on in this RC sample is on the order of a few percent. In this letter, we made use of the thirteenth data release version of the APOGEE RC catalogue described in \cite{Bovy2014}\footnote{Hence we call this the APOGEE RC sample the `Bovy' sample.} While this sample contains tens of thousands of RC stars, when we crossmatch this catalogue with \tgas\ and only select stars that have parallax precisions better than 30\% the total number of stars in this sample is only 639. 

The APO-1m sample, which is described in section 3 of \cite{Feuillet2016}, contains 324 potential RC stars and was generated using bright stars which have \hipparcos\ parallaxes better than 10\%. These stars were observed with the New Mexico State University (NMSU) 1-m telescope instead of the 2.5-m telescope used for the main APOGEE sample due to their brightness. The RC sample in this work was selected using the same procedures as in \cite{Bovy2014}. It is expected that the RGB contamination on in this RC sample is on the order of a few to 10 percent. We crossmatched this sample with \tgas\ and the various photometric surveys described above and further required that there was an estimated E$(B-V)_{i}$ from the dustmap of \cite{Green2015}. These cuts produce a final APO-1m sample of 218 RC stars. 

\subsection{APOKASC Red Clump Sample}
The APOGEE+Kepler (APOKASC) sample is comprised of stars in the Kepler field with seismic information that also have accompaning APOGEE spectra. The main distinction between the selection of this sample and those of the APOGEE RC sample is the use of asteroseismology. Namely, the stellar evolutionary status of these stars have seismically determined to be RC using the frequency period spacing \citep[e.g.][]{Pinsonneault2014, Elsworth2016}. This allows for a fairly clean selection of RC stars. Thus it is expected that the RGB contamination in this RC sample is very small (consistent with zero). As in Section~\ref{subsec:APOGEEsamp}, we only select stars with parallax precisions better than 30\% and with an E$(B-V)_{i}$ estimate from the dustmap of \cite{Green2015}  reducing the  APOKASC RC sample to 99 stars.

\subsection{Laney Red Clump Sample}
This sample is drawn from the crossmatch of \cite{Laney2012} and \tgas. Of the 226 brightest nearby bright RC stars in \cite{Laney2012}, 55 of them are found in \tgas. The reason for this is because, many of these stars are very bright and as such they were not reported in the first \gaia\ data release \citep{Gaia2016}. Many of these 55 stars have no reddening estimate from \cite{Green2015} and thus were removed from the sample. For this sample, we select to use the \tmass\ photometry rather than the more precise values from \cite{Laney2012} for consistency. We note however, since we properly account for the uncertainties in the photometry in our \hmodel, the use of either magnitudes do not affect the results. It is expected that the RGB contamination on in this RC sample is very small (probably consistent with zero). This sample adds an additional $\sim$20 RC stars to the final sample.

\section{The red clump model: A hierarchical approach} \label{sec:method}
\begin{figure}
	 \includegraphics[width=0.9\columnwidth]{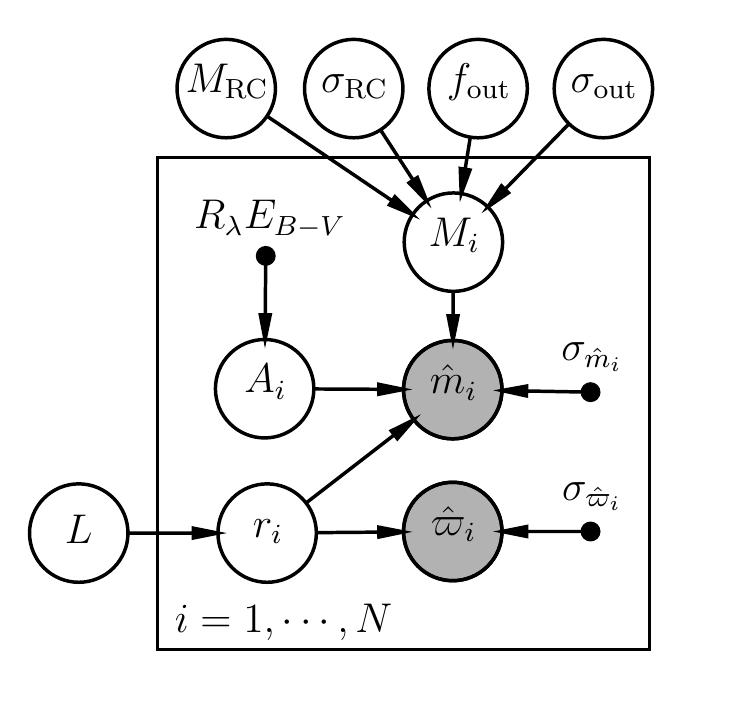}
	\caption{Illustrates the dependencies between the observed data and model parameters in the form of a probabilistic graphical model. Observed data is indicated by shaded circles while model parameters are denoted by open circles. Fixed parameters, such as the E($B-V$), the the extinction coefficient, R$_{\lambda}$, and uncertainties in data sources are represented by small filled black circles.   }
	\label{fig:PGM}
\end{figure}

Our approach to deriving the absolute magnitude and dispersion of the RC population in several bandpasses is a statistical one. We begin by modeling the absolute magnitude (in some bandpass) of the RC as a single Gaussian with an intrinsic magnitude, M$_{RC}$, with a dispersion, $\sigma_{\mathrm{M}_{RC}}$. The RC sample will likely have some contamination from the RGB, as these two features on the color-magnitude diagram overlap (for example see Figure~\ref{fig:CMD_RCall}). To account for this, we have added a second Gaussian `contamination' component with mean, M$_{out}$, and dispersion, $\sigma_{\mathrm{M}_{out}}$. with some contamination fraction which we denote as $f_\mathrm{out}$. For simplicity, we assume the contamination component has an intrinsic absolute magnitude centered at the same value as the RC population with a large dispersion (larger than 0.30~mag) that will be inferred (i.e. M$_{out}$~=~M$_{RC}$). We visualize the statistical model outlined in the section in the probabilistic graphical model (PGM) shown in Figure~\ref{fig:PGM}. 

In the context of our RC model, we want to compute the joint posterior probability distribution of our model given the observed data, i.e. $p(\theta_{\mathrm{RC}}, L, \alpha_{i} | \mathcal{D}_i)$,  where $\theta_{\mathrm{RC}}$ denotes the parameters of the RC model, i.e. $\theta_{\mathrm{RC}}$  = \{M$_{RC}$,  $\sigma_{\mathrm{M}_{RC}}$, M$_{out}$,  $\sigma_{\mathrm{M}_{out}}$, $f_\mathrm{out}$\},  L is the scale length of the distance prior (see Equation~\ref{eq:rprior}, below), and  $\mathcal{D}_i$ = $(\hat{m}_i$,  $\sigma_{\hat{m}_{i}}$, $\hat{\varpi}_{i}$, $\sigma_{\hat{\varpi}_{i}}$, E$(B-V)_{i})$ is the observed data. The $\alpha_{i}$~=~(r$_i$, A$_{i}$) represents the latent parameters, which include distance, r$_i$, and extinction, A$_{i}$, for every star that we will marginalize over.

Using Bayes' theorem, we can write down the (un-marginalized) posterior probability of the RC model as:
\begin{equation}
\begin{split}
&p(\theta_{\mathrm{RC}}, L, \{\alpha_i\}\ |\ \{\mathcal{D}_i \} ) \propto \\ &p(\theta_{\mathrm{RC}}, L) \prod_{i} p(\mathcal{D}_i\ |\ \theta_{\mathrm{RC}}, L, \alpha_{i}) \ p( \alpha_{i} | \theta_{\mathrm{RC}}, L ) ,
\end{split}
\end{equation}
where $p(\mathcal{D}_i |\ \theta_{\mathrm{RC}}, L, \alpha_{i}) $ is the object likelihood function and $p(\theta_{\mathrm{RC}}, L)$ is the  prior on the RC model parameters and  $p(\alpha_{i} | \theta_{\mathrm{RC}}, L )$ is the prior on the distance and extinction for each object.

The un-marginalized likelihood function per object is given as follows:
\begin{equation} 
p(\mathcal{D}_i |\ \theta_{\mathrm{RC}}, L, \alpha_{i}) \\= p(\hat{\varpi}_{i}\ |\ \varpi) \times\ p( \hat{m_i} |\ \theta_{\mathrm{RC}}, \alpha_{i}  ) ,
\label{eq:unmarglike}
\end{equation}
where 
\begin{equation}
p(\hat{\varpi}_{i}\ |\ r_i) = \mathcal{N}(\hat{\varpi}_{i}\ | 1/r_i, \sigma_{\hat{\varpi}_{i}})
\label{eq:parallax_like}
\end{equation}
and 
\begin{equation}
p(\hat{m}_i |\ \theta_{\mathrm{RC}},\alpha_{i} ) = \mathcal{N}(\hat{m}_i | m_i, \sigma_{\hat{m}_{i}})
\label{eq:phot_like}
\end{equation}

Where Equations~(\ref{eq:parallax_like}) and (\ref{eq:phot_like}) are the parallax and photometry likelihoods, respectively. In the above $ \mathcal{N}(x | \mu, \sigma^2)$ represents the normal or Gaussian probability distribution evaluated at x with mean, $\mu$, and dispersion, $\sigma$. The predicted parallax for each star is generated using a distance, r (which will be marginalized over), such that it is equal to $1/\mathrm{r}_i$.  In Equation~(\ref{eq:unmarglike}), the predicted apparent magnitude, $m_i$ is generated using the following:
\begin{equation}
m_i= \mathrm{M}_{i} + 5\log_{10}(r_i) - 5 + A_{i} , 
\end{equation}
where $A_{i}$ defines the extinction in the bandpass (which will be marginalized over) and M$_i$ is the predicted absolute magnitude which is drawn from:
\begin{equation}
\begin{split}
p(\mathrm{M}_i\ |\ \mathrm{M}_{RC} ,\sigma_{\mathrm{M}_{RC}},\sigma_{\mathrm{M}_{out}}, \mathrm{f}_{\mathrm{out}}) = \\ (1-f_\mathrm{out})\mathcal{N}(\mathrm{M}_i\ |\ \mathrm{M}_{RC},\sigma_{M_{RC}})\ + \\ f_\mathrm{out}\mathcal{N}(\mathrm{M}_i\ |\ \mathrm{M}_{RC},\sigma_{M_{out}}).
\end{split}
\label{eq:Mmodel}
\end{equation}
In Equation~(\ref{eq:Mmodel}), the absolute magnitude distribution is modeled with two Gaussian distributions: one for the RC population and one for the outlier population with a contamination fraction of $f_\mathrm{out}$.

Finally, we use generous uniform priors on all parameters except distance, extinction and the dispersion of the contamination population.
The prior used for distance is taken from \cite{Bailer-Jones2015}:
\begin{equation}
p(r_i\ |\ L) = \frac{1}{2L^3}r_i^2 \exp(-r_i/L),
\label{eq:rprior}
\end{equation}
where L  is both the scale-length of the distance prior and is also a hyperparameter (i.e. a parameterization of a prior on distance). Equation~\ref{eq:rprior} represents an exponentially decreasing space density prior. Since the scale-length will be inferred, as a hyperparameter, we place an additional uninformative uniform prior on L between 0.1 $<$ L $<$ 4 kpc. 
The prior used for the extinction is a Gaussian distribution centered on the value expected from the 3-d dustmap of \cite{Green2015} with a dispersion of 0.05 mag\footnote{The expected extinction is assumed as $A_{i} =$~R$_{\lambda}\ \times$~E$~(B-V)_i$.}.
 
The prior used for the the dispersion of the contamination population is a broad Gaussian distribution centered at 1.5~mag with a scale of 0.50~mag. We note here that the results are not significantly affected by changing this prior by more than 0.50 and 0.30~mag in the mean and scale, respectively, of the Gaussian prior. 

We made use of the most recent (version 2.12) python implementation of STAN code \citep{STAN2017} to draw samples from the sample the posterior distribution with different model parameters. We have used pySTAN with 10000 iterations and 5 chains with half of the iterations used for burn-in.

One significant advantage of modeling the RC with this \hmodel\ is that we are able to {\it infer} the distance and extinction for every star. This has allowed us to increase the sample size significantly because we include RC stars with with parallax precisions up to 30\% rather than the 10\% often used in the literature. However we note that our model is simplified and has many assumptions. For example, the parallaxes and photometry are assumed to be unbiased, and their errors thought to be correct and representative.

\section{Results and Discussion} \label{sec:result}
\begin{figure*}
	 \includegraphics[width=2.0\columnwidth]{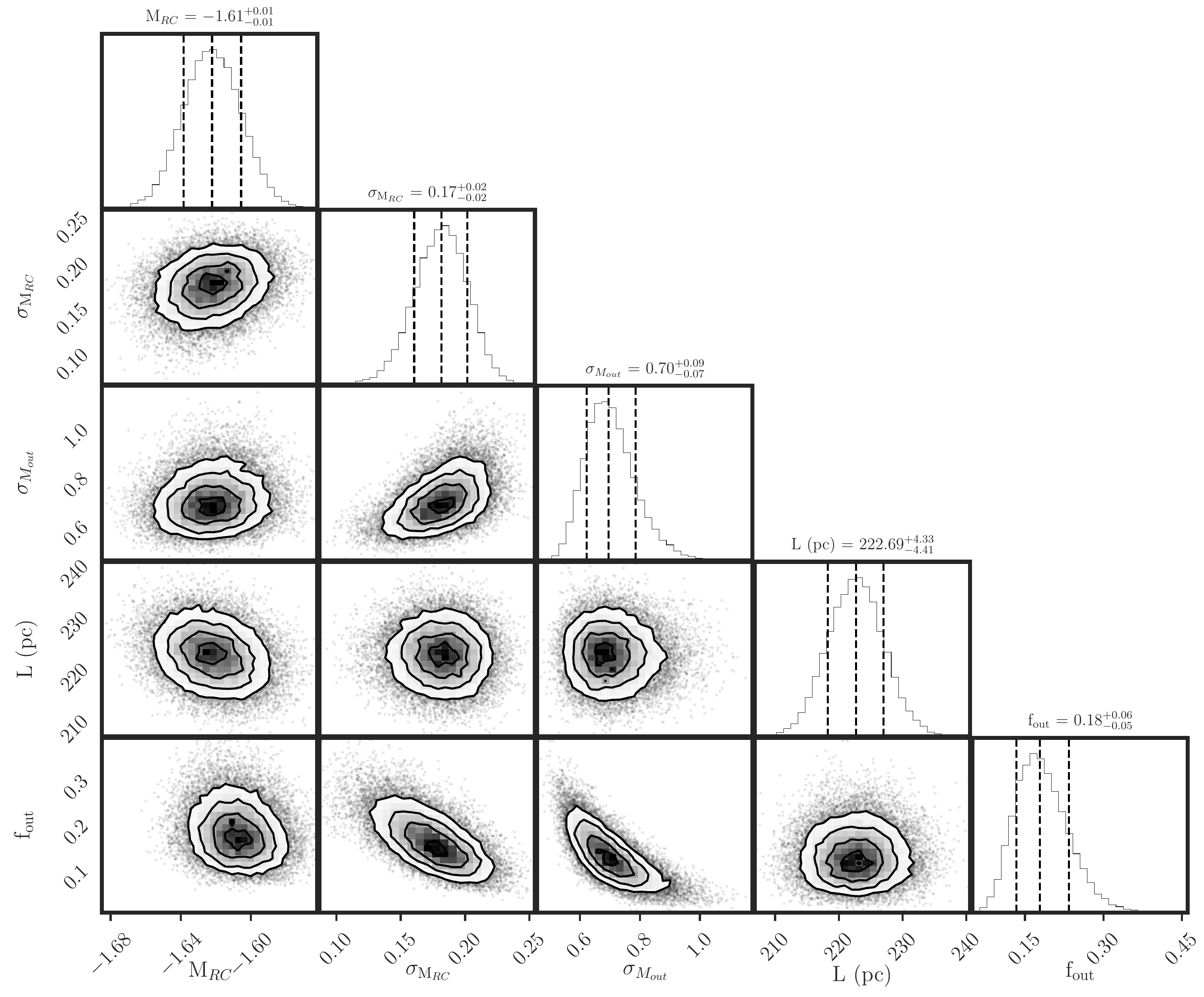}
	\caption{Corner plot of the displaying samples from the posterior probability for each of the parameters of the \hmodel\ for the RC sample in the \Ks\ band. }
		\label{fig:Kresult}
\end{figure*}
Figure~\ref{fig:Kresult} presents a corner plot of 25000 samples that have been drawn from the posterior distribution for each parameter of our RC model for the 2MASS \Ks\ bandpass. The extinction and distance for each star is not shown as these parameters are marginalized over. The 50th percentile, and 16th and 84th percentiles for each parameters are also displayed. While Figure~\ref{fig:Kresult} shows only the result for one of the eight bands studied (\Ks), Table~\ref{tab:results} presents the inferred values of the RC model for all eight bands. 

Now, we move on to describe the results in Table~\ref{tab:results}, and place them in the context of the literature, for each band.

\subsection{Gaia G}
In this work, we present a first measure of the intrinsic absolute magnitude and dispersion of the RC in the \gaia~G band. We have found that that the absolute magnitude in this wide-G band filter is M$_{\mathrm{G}}$~=~+0.44$\pm$0.01~mag with an intrinsic dispersion of 0.20$\pm$0.02~mag. This implies a floor distance precision of $\sim$10\% using this band. This result can be used to estimate the distance to all \gaia\ RC stars prior to future data releases at, conservatively, the 10\% level. We note that since the G band was not fully calibrated for DR1 \citep[e.g. see][]{Carrasco2016}, these results may change slightly with future releases.

\subsection{2MASS J, H, K}
The RC was first identified and calibrated in the 2MASS bands, specifically \Ks, by \cite{Alves2000}. The author found that the peak absolute magnitude of RC in \Ks\ was --1.62 $\pm$ 0.03. The same year \cite{Udalski2000} found a result that is consistent with this value. However, not long after these studies, \cite{vanHelshoecht2007} found that the RC has an absolute magnitude of --1.57 $\pm$ 0.05, while \cite{Groenewegen2008} found that the RC has an absolute magnitude of --1.54 $\pm$ 0.04 arguing that a selection bias towards bright stars caused the earlier results to be incorrect. More recently studies \citep[e.g.][]{Laney2012, Chen2017}, favour the brighter magnitude of --1.62 mag for the RC in the \Ks\ band. Our results are consistent with the more recent studies but inconsistent with the \cite{Groenewegen2008}. For example, most recently, \cite{Chen2017} found, using seismically determined RC stars from the Str\"{o}mgren survey for Asteroseismology and Galactic Archaeology, that the absolute magnitude of the RC is $-1.626 \pm 0.057$ in \Ks\ consistent with our results. 

Additionally, we point out that the \tmass\ band passes, more specifically \Ks, are often used to find RC stars because the absolute magnitude of the RC is likely to be only weakly dependent on age or metallicity \citep[e.g.][]{Udalski2000, vanHelshoecht2007, Groenewegen2008, Girardi2016}. 
In the J band, the absolute magnitude of RC in the \tmass\ J band can range from $\sim -0.92$ \citep{Bovy2017} to $-0.984 \pm$ 0.014 \citep{Laney2012}, to $-1.016 \pm 0.063$ \citep{Chen2017}. Our results are most consistent with those of \cite{Bovy2017} but are also in fair agreement with \cite{Chen2017} and \cite{Laney2012}. In the H band, the absolute magnitude of the RC can range in the literature with $-1.528 \pm 0.055$ found in \cite{Chen2017} to $-1.490 \pm$ 0.015 \citep{Laney2012}, both of which are in fair agreement with our value.

\subsection{WISE W1, W2, W3, W4}
The RC was first identified in \WISE\ bands W1 and W3 by \cite{Gokce2013}. They found that the absolute magnitude of the RC in W1 = $-1.64\pm0.03$ and in W3 = $-1.61\pm0.02$. More recently, \cite{Chen2017} found that the absolute magnitude of the RC in W1 = --1.69 $\pm$0.06, W2 = $-1.59 \pm$ 0.06, and W3 = $-1.752 \pm$ 0.06.  Our results in Table~\ref{tab:results} indicate the absolute magnitude of the RC in W1 is in good agreement with the values from \cite{Gokce2013} and \cite{Chen2017}. However, our results for W2 indicate the RC is brighter by 0.07 mag compared to \cite{Chen2017} with no estimate given in  \cite{Gokce2013}. In W3, our measurement of the absolute magnitude is  in between the values from \cite{Gokce2013} and \cite{Chen2017}. 

For many of the bandpasses above, we improve the precision with which we estimate the absolute magnitude of the RC, aided by the precise \tgas\ parallaxes and our \hmodel\ approach, where we properly model the uncertainties in the observables. We also expanded on the literature by estimating the intrinsic (de-noised) dispersion in W1, W2, and W3 of the RC and absolute magnitude in W4 bands for the first time. 

\begin{table*}
\caption{Red clump model parameters for the J, H, K, G, W1, W2, W3, and W4 bands}
\label{tab:results}
\centering 
\begin{tabular}{c c c c c c c c}
\hline\hline
Band& M$_{RC}$ & $\sigma_{\mathrm{M}_{RC}}$ & $\sigma_{\mathrm{M}_{out}}$  & L &  f$_{\mathrm{out}}$ & N & R$_{\lambda}$=$\frac{\mathrm{A}_{\lambda}}{\mathrm{E}(B-V)}$ \\
&(mag) & (mag) & (mag) & (pc)  & & & \\
\hline
G&+0.44$\pm$0.01 & 0.20 $\pm$0.02 & 0.75$\pm$0.08 & 215.6$\pm$4.2 & 0.18$\pm$0.04 & 972& 2.85\\
J &--0.93$\pm$0.01 & 0.20 $\pm$0.02 & 0.72$\pm$0.09 & 213.5$\pm$4.0 & 0.13$\pm$0.05 & 972 & 0.72\\
H&--1.46$\pm$0.01 & 0.17 $\pm$0.02 & 0.71$\pm$0.09 & 213.3$^{+4.1}_{-3.9}$ & 0.18$\pm$0.05 & 972 & 0.46\\
\Ks&--1.61$\pm$0.01 & 0.17 $\pm$0.02 & 0.70$^{+0.10}_{-0.08}$& 222.7$\pm$4.3 & 0.18$\pm0.05$&972 & 0.30\\
W1&--1.68$\pm$0.02 & 0.10 $\pm$0.04 & 0.73$^{+0.12}_{-0.09}$ & 231.5$\pm$4.8 & 0.15$\pm$0.04 & 936 & 0.18\\
W2&--1.69$\pm$0.02 & 0.20 $\pm$0.03 & 0.84$\pm$0.10 & 237.8$\pm$4.8 & 0.15$\pm$0.04 & 934 & 0.16\\
W3&--1.67$\pm$0.01 & 0.17 $\pm$0.02 & 0.74$\pm$0.08 & 228.3$\pm$4.6 & 0.18$\pm$0.05 & 936 & 0.16\\
W4&--1.76$\pm$0.01 & 0.16 $\pm$0.02 & 0.73$^{+0.09}_{-0.07}$ & 221.1$\pm$4.5 & 0.18$\pm$0.05 & 910 & 0.11\\
\hline  \hline     
\end{tabular}
\\
\raggedright
NOTE:  The bandpass is shown in column 1 while the absolute magnitude and dispersion in the absolute magnitude of the RC and `contaminate' population in that bandpass is listed in Columns 2, 3, 4, and 5, respectively. The inferred scale-length of the distance prior is tabulated in column 6 and the contaminate fraction, \fout\, can be found in column 7. The number of stars used in the inference and the assumed extinction coefficient for each band is tabulated in column 8 and 9, respectively. 
\end{table*}

\subsection{Error Shrinkage} \label{subsec:shrink}
\begin{figure}
	 \includegraphics[width=1.02\columnwidth]{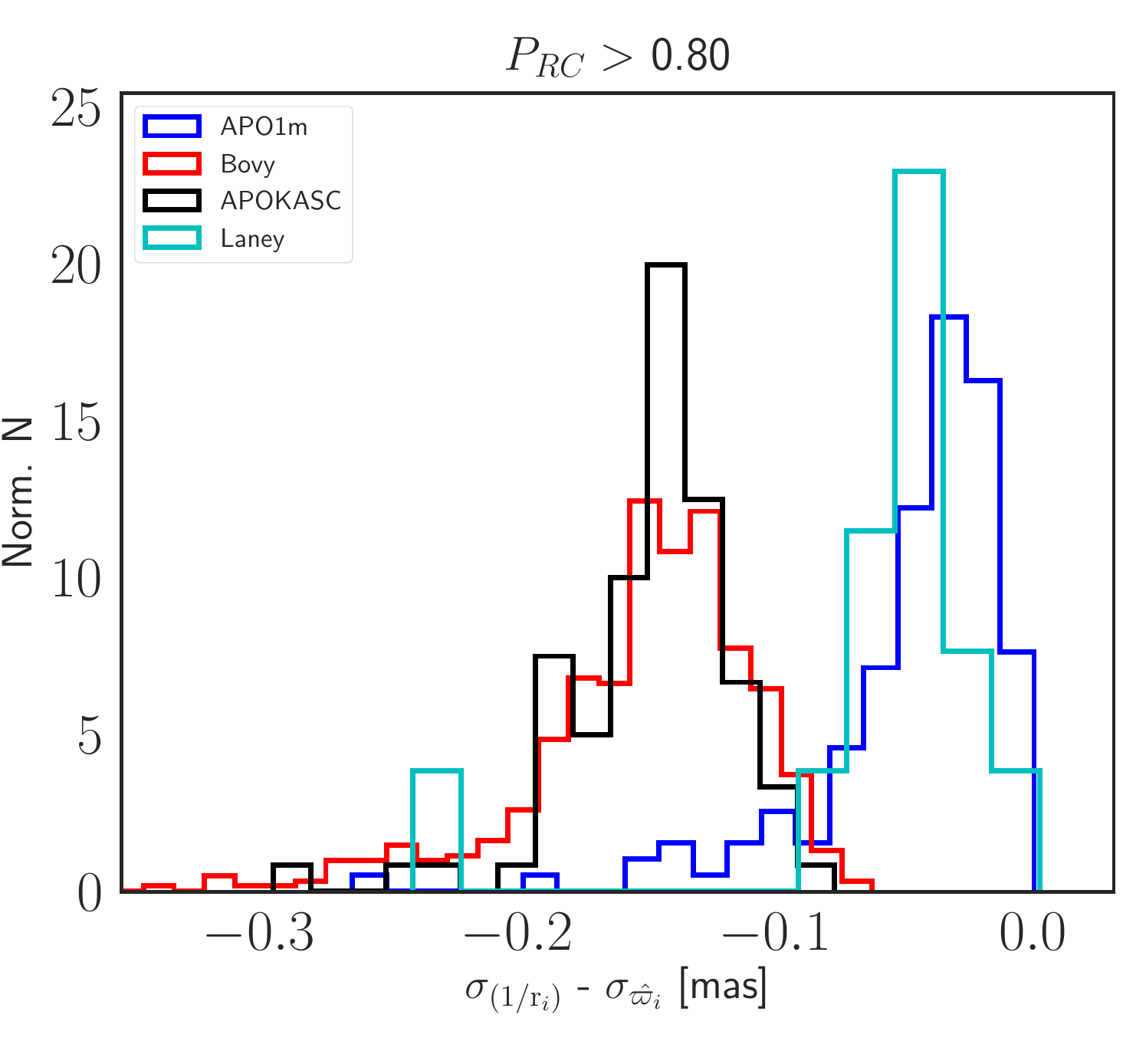}
	\caption{A histogram of the difference between the uncertainty in the inferred parallax from our \hmodel, $\sigma_{(1/\mathrm{r}_i)}$, and the \tgas\ parallax uncertainty, $\sigma_{\hat{\varpi_i}}$. The stars are split into their respective subsamples including APO1m (blue line), Bovy (red line), APOKASC (black line), Laney (cyan line). We only select those stars which have an inferred probability of belonging to the RC population that is larger than or equal to 80\%. The distributions illustrate that the inferred parallax precision is {\it better than} \tgas\ by $\sim$~0.15~mas, on average, for the distant APOKASC and Bovy subsamples and  0.03~mas for the more local APO1m and Laney subsamples. }
		\label{fig:error_shrink}
\end{figure}

One of the advantages of building a \hmodel\ to derive the absolute magnitude and dispersion of the RC, is that we properly model the uncertainties in the observables, allowing us to de-noise the parallax and magnitudes. This is illustrated in Figure~\ref{fig:error_shrink}, which shows the distribution of difference in the uncertainties in the inferred and \tgas\ parallaxes for probable\footnote{Probable RC stars are defined as those which have probabilities of being attributed to the RC component greater than or equal to 80\%. In this case, the probability for each star belonging to the RC is computed for every MCMC chain and the the median is taken.} RC stars are separated by subsample. The formal uncertainty in the inferred parallax from the \hmodel\ are always lower than the values quoted by \tgas. Distant subsamples (APOKASC and Bovy), the parallax precision of probable RC stars is {\it significantly better} compared to \tgas. The typical inferred parallax uncertainty is 0.15~mas lower than the \tgas\ uncertainty. For reference, we remind the reader that the median \tgas\ uncertainty in parallax for these samples are $\sim$0.30~mas. For the local subsamples (APO1m and Laney), where the \tgas\ precision is already very high, the uncertainties of the two are comparable. 

We note, that the decrease in the parallax uncertainty is not only a consequence of adopting a distance prior, but also from the hierarchical model itself. By inferring the properties of each star (e.g. its distance) and their subsequent population (e.g. the RC model parameters) simultaneously, each object borrows information from the others. As illustrated in \cite{Leistedt2017}, this is a classic ``shrinkage" property of hierarchical models 

\subsection{Possible Systematics}
\label{subsec:systematcs}
Furthermore, as a sanity check, we test (1) if different parallax cuts affect the result and (2) if the results on the absolute magnitude for every subsample are consistent with the final result. The outcome of the first sanity check was that the results are not affected by parallax cuts between 2--30\%. Below 2\% there are too few stars to achieve a reasonable estimate of the model parameters. Above 30\% there is no significant information gain for the RC model inference and the distances become more strongly determined by the prior.

The second test indicated that the inferred absolute magnitude of the RC for each subsample (described in section~\ref{sec:data}) are consistent, within the uncertainties, with those found in Table~\ref{tab:results}, except for the APOKASC subsample. In the APOKASC subsample containing 99 stars, the inferred absolute magnitude of the RC is $\sim$0.10~mag brighter in the \Ks\ band, for example compared to the results using the full combined 972 stars. This can be resolved by adding a $\sim$0.10~mas offset to the parallaxes in the APOKASC subsample (i.e. decreasing the distances to the APOKASC stars). Thus for the APOKASC sample, there is likely a non-zero systematic between the  parallaxes inferred in this work and \tgas.

We note that the second of these sanity checks was done because of the expected of systematics in the \tgas\ catalogue. These systematics are thought to be on the order of  $\pm$0.30~mas \citep[More specifically, there is a $\pm$0.10~mas systematic from potential global offset and $\pm$0.20~mas systematic which is both regional and color dependent,][]{Lindegren2016}, however the exact value has been a matter of debate. For example, \cite{Stassun2016} used eclipsing binary systems to show that there was a  typical parallax systematic,  $\varpi_{\mathrm{TGAS}} - \varpi_{\mathrm{EB}}$, on the order of --0.25~mas and as high as --0.39~mas in the field of the APOKASC subsample. This was consistent with the work of \cite{DeRidder2016} which indicated that there was a noticeable disagreement between \tgas\ and asteroseismic distances for a sample of 22 dwarf and subgiant solar-like oscillators and with \cite{Jao2016} using a sample of 612 single stars. Although, not long after, \cite{Davies2017} used a sample of nearly 850 asteroseismic giant stars, some of which are found in this work, and reported a systematic offset that was smaller than those of \cite{Stassun2016} but at least was found to go in the same direction (i.e. $\varpi_{\mathrm{TGAS}} - \varpi_{\mathrm{sesmic}} \sim$ -- 0.28~mas). However, more recent work using more than 100 RR Lyrae stars \citep{Sesar2017} and large numbers of asteroseismic targets \citep{Huber2017}, have either found no or very small parallax systematics that may be in tension with the results of \cite{Stassun2016} and \cite{Davies2017}. For reference, a systematic offset in the observed parallaxes by $\pm$0.10~mas will impact the inferred absolute magnitude of the RC by $\sim \pm$0.10~mag

\begin{figure}
	 \includegraphics[width=1.0\columnwidth]{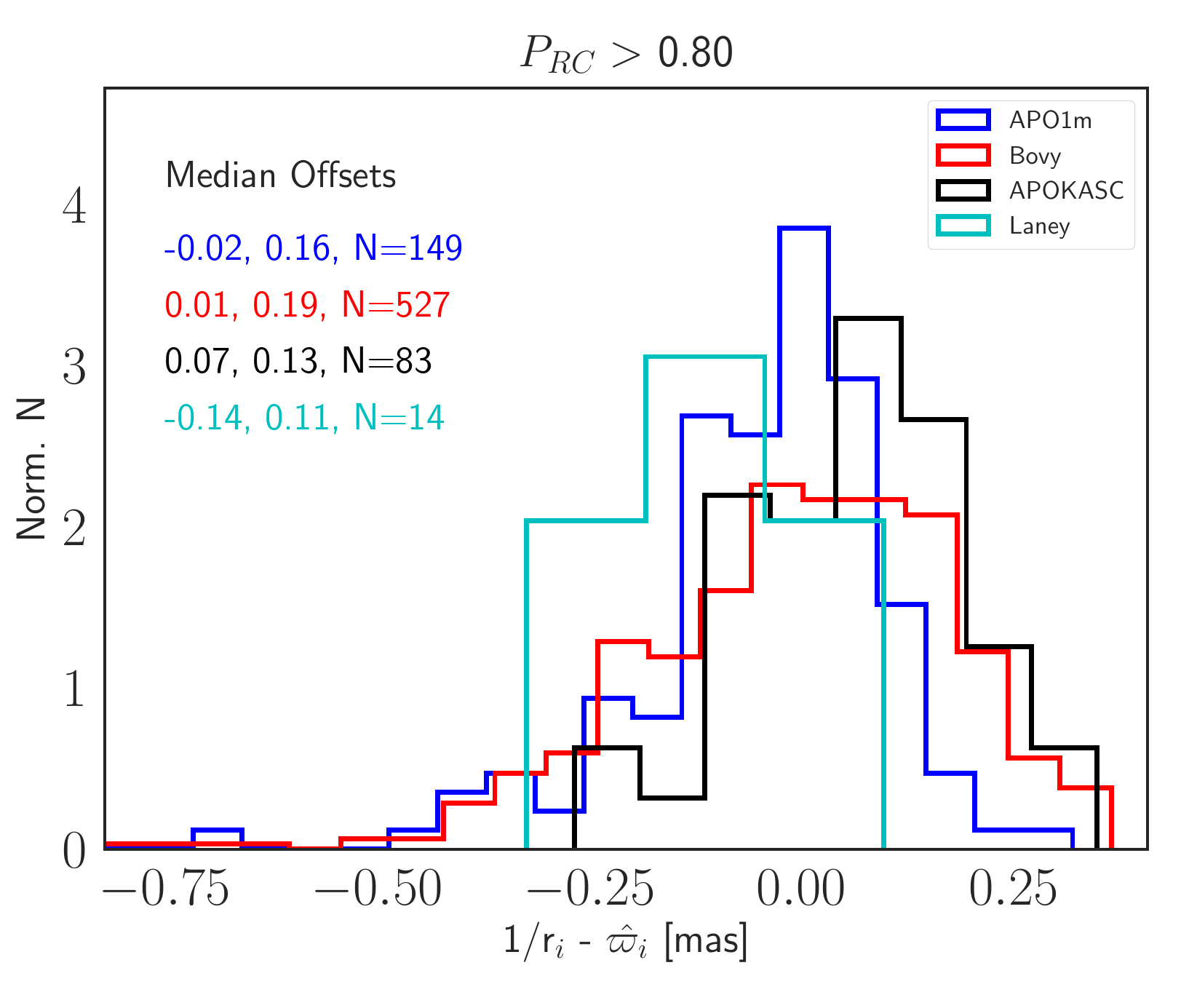}
	\caption{The normalized distribution of the difference between the inferred parallax from the \hmodel and the \tgas\ parallax. The line styles are the same as those in Figure~\ref{fig:error_shrink} and we only select those stars which have an inferred probability of belonging to the RC population that is larger than or equal to 80\%. The text on the left presents the median offset and dispersion with respect to \tgas\  for each sample, separately. The text is color-coded to match the legend. }
		\label{fig:sys}
\end{figure}

To further illustrate the parallax offsets found between this work and \tgas, in Figure~\ref{fig:sys} we plot the distribution of the difference between our inferred parallax for probable RC stars and the values from \tgas. As in Figure~\ref{fig:error_shrink}, the distributions are separated by subsample because each has a different distance range with the APO1m and Laney subsamples being nearby and the Bovy and APOKASC samples being much further. The median offset and its dispersion is also displayed in text on the right side of the histograms. Figure~\ref{fig:sys} indicates that for the APOKASC sample, there is a systematic offset whereby \tgas\ parallaxes are too large, by $\sim$0.10 mas, compared to our inferred parallaxes. This is consistent with the results of \cite{Huber2017}, which finds a much smaller parallax systematic in the Kepler field than both \cite{Stassun2016} and \cite{Davies2017}. For stars with the highest quality parallaxes, the systematic found by \cite{Davies2017} is consistent with our results. However, We note that the offset for the full combined sample is --0.003~mas with a dispersion of 0.24~mas. 

Therefore, for the full sample we do not include a systematic on the full all-sky sample, as suggested by \cite{Sesar2017}, on our final results as it would result in absolute magnitudes that are far to inconsistent with the literature. We also note that the final results do not significantly change by removing the APOKASC sample. We do point out that it may be unfair to compare the systematic offsets from different samples as it may be color, positional, and parallax dependent \citep[e.g.][]{Lindegren2016}. A fair comparison of the systematics in \tgas\ should be done only for stars of the same color, parallax distribution, and position on the sky. 

\section{Summary} \label{sec:summary}
In this work, we have put forward a first of its kind \hmodel\ to derive the intrinsic magnitude absolute magnitude and dispersion of helium burning RC stars. The \hmodel\ is advantageous because it fits for the properties of the RC sample while fully capturing many sources of uncertainties which are typically ignored including RGB contamination, dust, apparent magnitudes, and parallaxes. This new method has been applied to a sample of 972 RC stars which have \tgas\ parallaxes, with precisions better than 30\%, in order to update the zero-point absolute magnitude of the RC in eight photometric  bands including \tmass\ J, H, \Ks, \gaia\ G, and \WISE\ W1, W2, W3, and W4. We have also quantified the degree to which the RC should continue to be used as a standard candle by deriving the intrinsic dispersion of absolute magnitude (which will be small for the best standard candles).

We have shown that in every band studied the absolute magnitude in the RC has a dispersion that is less than $\sim$~0.20~mag and ranges from 0.10 -- 0.20~mag. Following standard error propagation, it can be shown that this yields a precision in distance of~$\sim$5--10\%. Additionally, we have updated the absolute magnitude of the RC in J, H, \Ks, W1, W2, and W3 and provide a first estimate in W4 and G. These values can be found in Table~\ref{tab:results}. Our results are in excellent to fair agreement with the literature. 

We also have looked at the offsets between the inferred parallaxes for stars in the work and \tgas. Consistent with most recent work \citep{Huber2017} we found a small offset between the inferred parallax and \tgas\ for the APOAKSC sample. This offset is much smaller than other results \citep[e.g.][]{Davies2017}. However, we remind the reader that these systematics are not trivial, and are color and positional dependent. 

One significant advantage to using the \hmodel\ for this work, as illustrated in Figure~\ref{fig:error_shrink}, is that not only can we quantify the absolute magnitude and dispersion of the RC, but we simultaneously infer more precise parallaxes for distant stars. This can be used in future releases to infer more precise distances than \gaia\ to distant RC stars.  

It has been shown in several studies \citep[e.g.][]{Girardi2001, Salaris2002, vanHelshoecht2007} that populations effects, specifically age and metallicity, binarity, and helium abundance, can alter the absolute magnitude of the RC. We do not yet control for these populations effects because either the age or metallicity is not known for a some stars in our sample. We also do not currently impose a prior for each star to have consistent colours. Therefore, it is likely that the dispersion in absolute magnitude of the RC will decrease when both the population effects and stellar colour are accounted for, thereby improving the quality of RC stars as standard candles. With the second release of the data  from the \gaia\ mission available soon, it is expected  that there will be significantly larger samples of RC stars, which will open the door for further investigation into these population effects with the methods outlined in this work.

\section*{Acknowledgements}
{\small 
K.H. is funded by the Simons Foundation Society of Fellows and the Flatiron Institute Center for Computational Astrophysics in New York City. KH would like to thank Johanna Coronado, Hans-Walter Rix, Iulia Simion, and Adrian Price-Whelan for conversations that enthusiastically contributed to this work. JB received support from the Natural Sciences and Engineering Research Council of Canada. BL was supported by NASA through the Einstein Postdoctoral Fellowship (award
number PF6-170154). JB also received partial support from an Alfred P. Sloan Fellowship and from the Simons Foundation. DWH was partially supported by the NSF (AST-1517237) and the Moore--Sloan Data Science Environment at NYU. 
This work has made use of the following python packages: Corner \citep{corner2016}, pySTAN, scipy, numpy, matplotlib, and astropy.
This project was developed in part at the 2016 NYC Gaia Sprint,
hosted by the Center for Computational Astrophysics at the
Simons Foundation in New York City. This work has made use
of data from the European Space Agency (ESA) mission Gaia
(http://www.cosmos.esa.int/gaia), processed by the Gaia Data
Processing and Analysis Consortium (DPAC; http://www.
cosmos.esa.int/web/gaia/dpac/consortium). Funding for the
DPAC has been provided by national institutions, in particular,
the institutions participating in the Gaia Multilateral Agreement.}

\bibliography{bibliography}

\begin{thebibliography}{47}
\expandafter\ifx\csname natexlab\endcsname\relax\def\natexlab#1{#1}\fi

\bibitem[{{Alves}(2000)}]{Alves2000}
{Alves} D.~R., 2000, \apj, 539, 732

\bibitem[{{Alves} {et~al}\mbox{.}(2002){Alves}, {Rejkuba}, {Minniti}, \&
  {Cook}}]{Alves2002}
{Alves} D.~R., {Rejkuba} M., {Minniti} D., {Cook} K.~H., 2002, \apjl, 573, L51

\bibitem[{{Astraatmadja} \& {Bailer-Jones}(2016)}]{Astraatmadja2016}
{Astraatmadja} T.~L., {Bailer-Jones} C.~A.~L., 2016, \apj, 832, 137

\bibitem[{{Bailer-Jones}(2015)}]{Bailer-Jones2015}
{Bailer-Jones} C.~A.~L., 2015, \pasp, 127, 994

\bibitem[{{Bovy}(2017)}]{Bovy2017}
{Bovy} J., 2017, \mnras, in press, arXiv:1704.05063

\bibitem[{{Bovy} {et~al}\mbox{.}(2014){Bovy}, {Nidever}, {Rix}, {Girardi},
  {Zasowski}, {Chojnowski}, {Holtzman}, {Epstein}, {Frinchaboy}, {Hayden},
  {Rodrigues}, {Majewski}, {Johnson}, {Pinsonneault}, {Stello}, {Allende
  Prieto}, {Andrews}, {Basu}, {Beers}, {Bizyaev}, {Burton}, {Chaplin}, {Cunha},
  {Elsworth}, {Garc{\'{\i}}a}, {Garc{\'{\i}}a-Her{\'n}andez}, {Garc{\'{\i}}a
  P{\'e}rez}, {Hearty}, {Hekker}, {Kallinger}, {Kinemuchi}, {Koesterke},
  {M{\'e}sz{\'a}ros}, {Mosser}, {O'Connell}, {Oravetz}, {Pan}, {Robin},
  {Schiavon}, {Schneider}, {Schultheis}, {Serenelli}, {Shetrone}, {Silva
  Aguirre}, {Simmons}, {Skrutskie}, {Smith}, {Stassun}, {Weinberg}, {Wilson},
  \& {Zamora}}]{Bovy2014}
{Bovy} J. {et~al.}, 2014, \apj, 790, 127

\bibitem[{Carpenter {et~al}\mbox{.}(2017)Carpenter, Gelman, Hoffman, Lee,
  Goodrich, Betancourt, Brubaker, Guo, Li, \& Riddell}]{STAN2017}
Carpenter B. {et~al.}, 2017, Journal of Statistical Software, 76, 1

\bibitem[{{Carrasco} {et~al}\mbox{.}(2016){Carrasco}, {Evans}, {Montegriffo},
  {Jordi}, {van Leeuwen}, {Riello}, {Voss}, {De Angeli}, {Busso}, {Fabricius},
  {Cacciari}, {Weiler}, {Pancino}, {Brown}, {Holland}, {Burgess}, {Osborne},
  {Altavilla}, {Gebran}, {Ragaini}, {Galleti}, {Cocozza}, {Marinoni},
  {Bellazzini}, {Bragaglia}, {Federici}, \&
  {Balaguer-N{\'u}{\~n}ez}}]{Carrasco2016}
{Carrasco} J.~M. {et~al.}, 2016, \aap, 595, A7

\bibitem[{{Chen} {et~al}\mbox{.}(2017){Chen}, {Casagrande}, {Zhao}, {Bovy},
  {Silva Aguirre}, {Zhao}, \& {Jia}}]{Chen2017}
{Chen} Y.~Q., {Casagrande} L., {Zhao} G., {Bovy} J., {Silva Aguirre} V., {Zhao}
  J.~K., {Jia} Y.~P., 2017, ArXiv e-prints:1704.03903

\bibitem[{{Cutri} {et~al}\mbox{.}(2003){Cutri}, {Skrutskie}, {van Dyk},
  {Beichman}, {Carpenter}, {Chester}, {Cambresy}, {Evans}, {Fowler}, {Gizis},
  {Howard}, {Huchra}, {Jarrett}, {Kopan}, {Kirkpatrick}, {Light}, {Marsh},
  {McCallon}, {Schneider}, {Stiening}, {Sykes}, {Weinberg}, {Wheaton},
  {Wheelock}, \& {Zacarias}}]{Cutri2003}
{Cutri} R.~M. {et~al.}, 2003, VizieR Online Data Catalog, 2246, 0

\bibitem[{{Davies} {et~al}\mbox{.}(2017){Davies}, {Lund}, {Miglio}, {Elsworth},
  {Kuszlewicz}, {North}, {Rendle}, {Chaplin}, {Rodrigues}, {Campante},
  {Girardi}, {Hale}, {Hall}, {Jones}, {Kawaler}, {Roxburgh}, \&
  {Schofield}}]{Davies2017}
{Davies} G.~R. {et~al.}, 2017, \aap, 598, L4

\bibitem[{{De Ridder} {et~al}\mbox{.}(2016){De Ridder}, {Molenberghs}, {Eyer},
  \& {Aerts}}]{DeRidder2016}
{De Ridder} J., {Molenberghs} G., {Eyer} L., {Aerts} C., 2016, \aap, 595, L3

\bibitem[{{Elsworth} {et~al}\mbox{.}(2016){Elsworth}, {Hekker}, {Basu}, \&
  {Davies}}]{Elsworth2016}
{Elsworth} Y., {Hekker} S., {Basu} S., {Davies} G., 2016, arXiv:1612.04751

\bibitem[{{Feuillet} {et~al}\mbox{.}(2016){Feuillet}, {Bovy}, {Holtzman},
  {Girardi}, {MacDonald}, {Majewski}, \& {Nidever}}]{Feuillet2016}
{Feuillet} D.~K., {Bovy} J., {Holtzman} J., {Girardi} L., {MacDonald} N.,
  {Majewski} S.~R., {Nidever} D.~L., 2016, \apj, 817, 40

\bibitem[{Foreman-Mackey(2016)}]{corner2016}
Foreman-Mackey D., 2016, The Journal of Open Source Software, 24

\bibitem[{{Gaia Collaboration} {et~al}\mbox{.}(2016){Gaia Collaboration},
  {Brown}, {Vallenari}, {Prusti}, {de Bruijne}, {Mignard}, {Drimmel},
  {Babusiaux}, {Bailer-Jones}, {Bastian}, \& et~al.}]{Gaia2016}
{Gaia Collaboration} {et~al.}, 2016, \aap, 595, A2

\bibitem[{{Girardi}(2016)}]{Girardi2016}
{Girardi} L., 2016, \araa, 54, 95

\bibitem[{{Girardi} \& {Salaris}(2001)}]{Girardi2001}
{Girardi} L., {Salaris} M., 2001, \mnras, 323, 109

\bibitem[{{Green} {et~al}\mbox{.}(2015){Green}, {Schlafly}, {Finkbeiner},
  {Rix}, {Martin}, {Burgett}, {Draper}, {Flewelling}, {Hodapp}, {Kaiser},
  {Kudritzki}, {Magnier}, {Metcalfe}, {Price}, {Tonry}, \&
  {Wainscoat}}]{Green2015}
{Green} G.~M. {et~al.}, 2015, \apj, 810, 25

\bibitem[{{Groenewegen}(2008)}]{Groenewegen2008}
{Groenewegen} M.~A.~T., 2008, \aap, 488, 935

\bibitem[{{Hawkins} {et~al}\mbox{.}(2016){Hawkins}, {Masseron}, {Jofr{\'e}},
  {Gilmore}, {Elsworth}, \& {Hekker}}]{Hawkins2016b}
{Hawkins} K., {Masseron} T., {Jofr{\'e}} P., {Gilmore} G., {Elsworth} Y.,
  {Hekker} S., 2016, \aap, 594, A43

\bibitem[{{Holtzman} {et~al}\mbox{.}(2015){Holtzman}, {Shetrone}, {Johnson},
  {Allende Prieto}, {Anders}, {Andrews}, {Beers}, {Bizyaev}, {Blanton}, {Bovy},
  {Carrera}, {Chojnowski}, {Cunha}, {Eisenstein}, {Feuillet}, {Frinchaboy},
  {Galbraith-Frew}, {Garc{\'{\i}}a P{\'e}rez}, {Garc{\'{\i}}a-Hern{\'a}ndez},
  {Hasselquist}, {Hayden}, {Hearty}, {Ivans}, {Majewski}, {Martell},
  {Meszaros}, {Muna}, {Nidever}, {Nguyen}, {O'Connell}, {Pan}, {Pinsonneault},
  {Robin}, {Schiavon}, {Shane}, {Sobeck}, {Smith}, {Troup}, {Weinberg},
  {Wilson}, {Wood-Vasey}, {Zamora}, \& {Zasowski}}]{Holtzman2015}
{Holtzman} J.~A. {et~al.}, 2015, \aj, 150, 148

\bibitem[{{Huber} {et~al}\mbox{.}(2017){Huber}, {Zinn}, {Bojsen-Hansen},
  {Pinsonneault}, {Sahlholdt}, {Serenelli}, {Silva Aguirre}, {Stassun},
  {Stello}, {Tayar}, {Bastien}, {Bedding}, {Buchhave}, {Chaplin}, {Davies},
  {Garcia}, {Latham}, {Mathur}, {Mosser}, \& {Sharma}}]{Huber2017}
{Huber} D. {et~al.}, 2017, ArXiv e-prints:1705.04697

\bibitem[{{Jao} {et~al}\mbox{.}(2016){Jao}, {Henry}, {Riedel}, {Winters},
  {Slatten}, \& {Gies}}]{Jao2016}
{Jao} W.-C., {Henry} T.~J., {Riedel} A.~R., {Winters} J.~G., {Slatten} K.~J.,
  {Gies} D.~R., 2016, \apjl, 832, L18

\bibitem[{{Jordi} {et~al}\mbox{.}(2010){Jordi}, {Gebran}, {Carrasco}, {de
  Bruijne}, {Voss}, {Fabricius}, {Knude}, {Vallenari}, {Kohley}, \&
  {Mora}}]{Jordi2010}
{Jordi} C. {et~al.}, 2010, \aap, 523, A48

\bibitem[{{Laney}, {Joner} \& {Pietrzy{\'n}ski}(2012){Laney}, {Joner}, \&
  {Pietrzy{\'n}ski}}]{Laney2012}
{Laney} C.~D., {Joner} M.~D., {Pietrzy{\'n}ski} G., 2012, \mnras, 419, 1637

\bibitem[{{Leistedt} \& {Hogg}(2017)}]{Leistedt2017}
{Leistedt} B., {Hogg} D.~W., 2017, ArXiv e-prints:1703.08112

\bibitem[{{Lindegren} {et~al}\mbox{.}(2016){Lindegren}, {Lammers}, {Bastian},
  {Hern{\'a}ndez}, {Klioner}, {Hobbs}, {Bombrun}, {Michalik}, {Ramos-Lerate},
  {Butkevich}, {Comoretto}, {Joliet}, {Holl}, {Hutton}, {Parsons},
  {Steidelm{\"u}ller}, {Abbas}, {Altmann}, {Andrei}, {Anton}, {Bach},
  {Barache}, {Becciani}, {Berthier}, {Bianchi}, {Biermann}, {Bouquillon},
  {Bourda}, {Br{\"u}semeister}, {Bucciarelli}, {Busonero}, {Carlucci},
  {Casta{\~n}eda}, {Charlot}, {Clotet}, {Crosta}, {Davidson}, {de Felice},
  {Drimmel}, {Fabricius}, {Fienga}, {Figueras}, {Fraile}, {Gai}, {Garralda},
  {Geyer}, {Gonz{\'a}lez-Vidal}, {Guerra}, {Hambly}, {Hauser}, {Jordan},
  {Lattanzi}, {Lenhardt}, {Liao}, {L{\"o}ffler}, {McMillan}, {Mignard}, {Mora},
  {Morbidelli}, {Portell}, {Riva}, {Sarasso}, {Serraller}, {Siddiqui}, {Smart},
  {Spagna}, {Stampa}, {Steele}, {Taris}, {Torra}, {van Reeven}, {Vecchiato},
  {Zschocke}, {de Bruijne}, {Gracia}, {Raison}, {Lister}, {Marchant},
  {Messineo}, {Soffel}, {Osorio}, {de Torres}, \& {O'Mullane}}]{Lindegren2016}
{Lindegren} L. {et~al.}, 2016, \aap, 595, A4

\bibitem[{{Majewski} {et~al}\mbox{.}(2015){Majewski}, {Schiavon}, {Frinchaboy},
  {Allende Prieto}, {Barkhouser}, {Bizyaev}, {Blank}, {Brunner}, {Burton},
  {Carrera}, {Chojnowski}, {Cunha}, {Epstein}, {Fitzgerald}, {Garcia Perez},
  {Hearty}, {Henderson}, {Holtzman}, {Johnson}, {Lam}, {Lawler}, {Maseman},
  {Meszaros}, {Nelson}, {Coung Nguyen}, {Nidever}, {Pinsonneault}, {Shetrone},
  {Smee}, {Smith}, {Stolberg}, {Skrutskie}, {Walker}, {Wilson}, {Zasowski},
  {Anders}, {Basu}, {Beland}, {Blanton}, {Bovy}, {Brownstein}, {Carlberg},
  {Chaplin}, {Chiappini}, {Eisenstein}, {Elsworth}, {Feuillet}, {Fleming},
  {Galbraith-Frew}, {Garcia}, {Anibal Garcia-Hernandez}, {Gillespie},
  {Girardi}, {Gunn}, {Hasselquist}, {Hayden}, {Hekker}, {Ivans}, {Kinemuchi},
  {Klaene}, {Mahadevan}, {Mathur}, {Mosser}, {Muna}, {Munn}, {Nichol},
  {O'Connell}, {Robin}, {Rocha-Pinto}, {Schultheis}, {Serenelli}, {Shane},
  {Silva Aguirre}, {Sobeck}, {Thompson}, {Troup}, {Weinberg}, \&
  {Zamora}}]{Majewski2015}
{Majewski} S.~R. {et~al.}, 2015, ArXiv e-prints:1509.05420

\bibitem[{{McWilliam} \& {Zoccali}(2010)}]{McWilliam2010}
{McWilliam} A., {Zoccali} M., 2010, \apj, 724, 1491

\bibitem[{{Michalik}, {Lindegren} \& {Hobbs}(2015){Michalik}, {Lindegren}, \&
  {Hobbs}}]{Michalik2015}
{Michalik} D., {Lindegren} L., {Hobbs} D., 2015, \aap, 574, A115

\bibitem[{{Ness} {et~al}\mbox{.}(2015){Ness}, {Hogg}, {Rix}, {Ho}, \&
  {Zasowski}}]{Ness2015}
{Ness} M., {Hogg} D.~W., {Rix} H.-W., {Ho} A.~Y.~Q., {Zasowski} G., 2015, \apj,
  808, 16

\bibitem[{{Nidever} {et~al}\mbox{.}(2014){Nidever}, {Bovy}, {Bird}, {Andrews},
  {Hayden}, {Holtzman}, {Majewski}, {Smith}, {Robin}, {Garc{\'{\i}}a
  P{\'e}rez}, {Cunha}, {Allende Prieto}, {Zasowski}, {Schiavon}, {Johnson},
  {Weinberg}, {Feuillet}, {Schneider}, {Shetrone}, {Sobeck},
  {Garc{\'{\i}}a-Hern{\'a}ndez}, {Zamora}, {Rix}, {Beers}, {Wilson},
  {O'Connell}, {Minchev}, {Chiappini}, {Anders}, {Bizyaev}, {Brewington},
  {Ebelke}, {Frinchaboy}, {Ge}, {Kinemuchi}, {Malanushenko}, {Malanushenko},
  {Marchante}, {M{\'e}sz{\'a}ros}, {Oravetz}, {Pan}, {Simmons}, \&
  {Skrutskie}}]{Nidever2014}
{Nidever} D.~L. {et~al.}, 2014, \apj, 796, 38

\bibitem[{{Paczy{\'n}ski} \& {Stanek}(1998)}]{Paczyski1998}
{Paczy{\'n}ski} B., {Stanek} K.~Z., 1998, \apjl, 494, L219

\bibitem[{{Pinsonneault} {et~al}\mbox{.}(2014){Pinsonneault}, {Elsworth},
  {Epstein}, {Hekker}, {M{\'e}sz{\'a}ros}, {Chaplin}, {Johnson},
  {Garc{\'{\i}}a}, {Holtzman}, {Mathur}, {Garc{\'{\i}}a P{\'e}rez}, {Silva
  Aguirre}, {Girardi}, {Basu}, {Shetrone}, {Stello}, {Allende Prieto}, {An},
  {Beck}, {Beers}, {Bizyaev}, {Bloemen}, {Bovy}, {Cunha}, {De Ridder},
  {Frinchaboy}, {Garc{\'{\i}}a-Hern{\'a}ndez}, {Gilliland}, {Harding},
  {Hearty}, {Huber}, {Ivans}, {Kallinger}, {Majewski}, {Metcalfe}, {Miglio},
  {Mosser}, {Muna}, {Nidever}, {Schneider}, {Serenelli}, {Smith}, {Tayar},
  {Zamora}, \& {Zasowski}}]{Pinsonneault2014}
{Pinsonneault} M.~H. {et~al.}, 2014, \apjs, 215, 19

\bibitem[{{Salaris} \& {Girardi}(2002)}]{Salaris2002}
{Salaris} M., {Girardi} L., 2002, \mnras, 337, 332

\bibitem[{{Sesar} {et~al}\mbox{.}(2017){Sesar}, {Fouesneau}, {Price-Whelan},
  {Bailer-Jones}, {Gould}, \& {Rix}}]{Sesar2017}
{Sesar} B., {Fouesneau} M., {Price-Whelan} A.~M., {Bailer-Jones} C.~A.~L.,
  {Gould} A., {Rix} H.-W., 2017, \apj, 838, 107

\bibitem[{{Stanek}, {Zaritsky} \& {Harris}(1998){Stanek}, {Zaritsky}, \&
  {Harris}}]{Stanek1998}
{Stanek} K.~Z., {Zaritsky} D., {Harris} J., 1998, \apjl, 500, L141

\bibitem[{{Stassun} \& {Torres}(2016)}]{Stassun2016}
{Stassun} K.~G., {Torres} G., 2016, \apjl, 831, L6

\bibitem[{{Udalski}(2000)}]{Udalski2000}
{Udalski} A., 2000, \apjl, 531, L25

\bibitem[{{Udalski} {et~al}\mbox{.}(1998){Udalski}, {Szymanski}, {Kubiak},
  {Pietrzynski}, {Wozniak}, \& {Zebrun}}]{Udalski1998}
{Udalski} A., {Szymanski} M., {Kubiak} M., {Pietrzynski} G., {Wozniak} P.,
  {Zebrun} K., 1998, \actaa, 48, 1

\bibitem[{{van Helshoecht} \& {Groenewegen}(2007)}]{vanHelshoecht2007}
{van Helshoecht} V., {Groenewegen} M.~A.~T., 2007, \aap, 463, 559

\bibitem[{{van Leeuwen} {et~al}\mbox{.}(2017){van Leeuwen}, {Evans}, {De
  Angeli}, {Jordi}, {Busso}, {Cacciari}, {Riello}, {Pancino}, {Altavilla},
  {Brown}, {Burgess}, {Carrasco}, {Cocozza}, {Cowell}, {Davidson}, {De Luise},
  {Fabricius}, {Galleti}, {Gilmore}, {Giuffrida}, {Hambly}, {Harrison},
  {Hodgkin}, {Holland}, {MacDonald}, {Marinoni}, {Montegriffo}, {Osborne},
  {Ragaini}, {Richards}, {Rowell}, {Voss}, {Walton}, {Weiler}, {Castellani},
  {Delgado}, {H{\o}g}, {van Leeuwen}, {Millar}, {Pagani}, {Piersimoni},
  {Pulone}, {Rixon}, {Suess}, {Wyrzykowski}, {Yoldas}, {Alecu}, {Allan},
  {Balaguer-N{\'u}{\~n}ez}, {Barstow}, {Bellazzini}, {Belokurov},
  {Blagorodnova}, {Bonfigli}, {Bragaglia}, {Brown}, {Bunclark}, {Buonanno},
  {Burgon}, {Campbell}, {Collins}, {Cross}, {Ducourant}, {van Elteren},
  {Evans}, {Federici}, {Fern{\'a}ndez-Hern{\'a}ndez}, {Figueras}, {Fraser},
  {Fyfe}, {Gebran}, {Heyrovsky}, {Holl}, {Holland}, {Iannicola}, {Irwin},
  {Koposov}, {Krone-Martins}, {Mann}, {Marrese}, {Masana}, {Munari}, {Ortiz},
  {Ouzounis}, {Peltzer}, {Portell}, {Read}, {Terrett}, {Torra}, {Trager},
  {Troisi}, {Valentini}, {Vallenari}, \& {Wevers}}]{vanLeeuwen2017}
{van Leeuwen} F. {et~al.}, 2017, \aap, 599, A32

\bibitem[{{Wright} {et~al}\mbox{.}(2010){Wright}, {Eisenhardt}, {Mainzer},
  {Ressler}, {Cutri}, {Jarrett}, {Kirkpatrick}, {Padgett}, {McMillan},
  {Skrutskie}, {Stanford}, {Cohen}, {Walker}, {Mather}, {Leisawitz}, {Gautier},
  {McLean}, {Benford}, {Lonsdale}, {Blain}, {Mendez}, {Irace}, {Duval}, {Liu},
  {Royer}, {Heinrichsen}, {Howard}, {Shannon}, {Kendall}, {Walsh}, {Larsen},
  {Cardon}, {Schick}, {Schwalm}, {Abid}, {Fabinsky}, {Naes}, \&
  {Tsai}}]{Wright2010}
{Wright} E.~L. {et~al.}, 2010, \aj, 140, 1868

\bibitem[{{Xue} {et~al}\mbox{.}(2016){Xue}, {Jiang}, {Gao}, {Liu}, {Wang}, \&
  {Li}}]{Xue2016}
{Xue} M., {Jiang} B.~W., {Gao} J., {Liu} J., {Wang} S., {Li} A., 2016, \apjs,
  224, 23

\bibitem[{{Yaz G{\"o}k{\c c}e} {et~al}\mbox{.}(2013){Yaz G{\"o}k{\c c}e},
  {Bilir}, {{\"O}zt{\"u}rkmen}, {Duran}, {Ak}, {Ak}, \& {Karaali}}]{Gokce2013}
{Yaz G{\"o}k{\c c}e} E., {Bilir} S., {{\"O}zt{\"u}rkmen} N.~D., {Duran} {\c
  S}., {Ak} T., {Ak} S., {Karaali} S., 2013, \na, 25, 19

\bibitem[{{Yuan}, {Liu} \& {Xiang}(2013){Yuan}, {Liu}, \& {Xiang}}]{Yuan2013}
{Yuan} H.~B., {Liu} X.~W., {Xiang} M.~S., 2013, \mnras, 430, 2188

\end{thebibliography}
\label{lastpage}

\end{document}